\documentstyle[aps,preprint]{revtex}

\begin{document}

\draft

%\twocolumn

%\tighten

\hyphenation{}

\title{Theory of small-polaron band conduction in ultrapure organic crystals}

\author{K. Hannewald and P. A. Bobbert}

\address{Group Polymer Physics, Eindhoven Polymer Laboratories, Technische 
Universiteit Eindhoven, P.O.Box 513, 5600 MB Eindhoven, The Netherlands}

\date{\today}

\maketitle

\begin{abstract}

We present a novel theory of charge-carrier mobilities in organic molecular 
crystals of high purity. Our approach is based on Holstein's original concept 
of small-polaron bands but generalized with respect to the inclusion of 
nonlocal electron-phonon coupling. We derive an explicit expression for the 
mobilities as a function of temperature and, using {\it ab-initio} methods to 
obtain the material parameters, we demonstrate its predictive 
power by applying it to naphthalene. The results show a remarkably good 
agreement with experiments and provide new insight into the difference 
between electron and hole mobilities as well as their peculiar algebraic 
and anisotropic temperature dependences.

\end{abstract}


\narrowtext

Organic semiconductors with $\pi$-conjugation are very promising 
materials for low-cost and easy-to-process electronic and optoelectronic 
devices such as light-emitting diodes \cite{Burroughes,Friend}, lasers 
\cite{Tessler,Vardeny}, solar cells \cite{Sariciftci,Granstroem}, 
and thin-film transistors 
\cite{Garnier,Dodabalapur,Schoonveld,Stutzmann,Hepp}. 
Besides the polymeric materials, a very  
important class of organic semiconductors are molecular 
crystals of high purity. Due to their high degree of structural order, such 
crystals are ideal candidates for the investigation of the 
{\it intrinsic} excitations and charge-carrier transport phenomena in 
organic solids. Consequently, many interesting fundamental 
experimental and theoretical studies have been performed in recent years and 
new results emerge at a rapid pace
\cite{Coropceanu,Northrup,Venuti,deWijs,Cheng}. 
A particularly important and challenging topic is the 
understanding of the charge-carrier mobilities in these crystals, 
as the mobility is a fundamental 
material property and a central quantity for 
optimization of device performance. 

In comparison to covalently bonded inorganic semiconductors, organic molecular 
crystals exhibit weak intermolecular bonds and, hence, 
narrower electron bands and stronger electron-lattice interaction. As a 
result, the dressing of the charge carriers by phonon clouds, i.e., 
polaron formation, becomes quite important. Pioneering theoretical work in 
this area was done by Holstein \cite{Holstein} who introduced the concept of 
small-polaron bands for a {\it local} (on-site) electron-phonon 
coupling. The main 
prediction of his theory is that the mobilities initially decrease with 
rising temperature $T$ due to bandwidth narrowing 
but increase again for higher $T$ due to phonon-assisted hopping. 
Two decades after Holstein's prediction, the interplay between metallic 
(bandlike) conduction and activated (hopping) transport in organic crystals 
was reported for the electrons in naphthalene but not for the holes 
\cite{Schein,Karl,Warta}. Later, Kenkre {\it et al.} 
were able to fit the measured electron mobilities reasonably well 
to Holstein's model, assuming directionally-dependent local-coupling constants 
\cite{Kenkre}. Despite the success of such a {\it fitting} procedure, the 
lack of a first-principles description of charge-carrier mobilities in 
organic crystals has left several fundamental questions unanswered. 
This concerns, in particular, the different behavior of electrons and holes, 
the microscopic origin of the crystallographic anisotropy in the $T$ 
dependence, and the influence of {\it nonlocal} (Peierls-type) couplings 
such as present in the Su-Schrieffer-Heeger model \cite{Su}. 
Moreover, the fundamental question has been raised whether or not the 
concept of small-polaron bands is applicable at all, since band motion 
may be washed out by even weak disorder \cite{Emin}.

In this Letter, we answer the above questions by developing a novel 
first-principles theory of charge-carrier mobilities in organic crystals. 
We present strong evidence 
for the usefulness of the small-polaron band picture once Holstein's 
original model is extended towards the inclusion of nonlocal 
electron-phonon coupling. While a few theoretical studies on nonlocal 
coupling have been performed in the past 
\cite{Friedman,Gosar,Sumi,Munn,Zhao}, it was only recently that the 
importance of these contributions could be demonstrated 
explicitly in the case of polaron bandwidth narrowing in oligo-acene crystals 
\cite{Hannewald}. Here, by means of the Kubo formalism, we obtain an 
explicit expression for the mobilities, including both local {\it and} 
nonlocal coupling. The practical usefulness of this approach is then 
exemplified by {\it ab-initio} studies for naphthalene 
(${\rm C}_{10}{\rm H}_8$) crystals.

We consider a mixed Holstein-Peierls model for the interaction between 
electrons (holes) and phonons. In a tight-binding description, this 
corresponds to a Hamiltonian of the form 
$H=H_{\rm el}+H_{\rm ph}+H_{\rm el-ph}$ 
where 
\begin{eqnarray} \label{H_EL}
H_{\rm el} &=& \sum_{mn} \varepsilon_{mn} a_m^{\dag} a_n  \,\,, 
\\ \label{H_PH} 
H_{\rm ph} &=& \sum_{{\bf Q}} \hbar \omega_{\bf Q} 
(b_{\bf Q}^{\dag} b_{\bf Q}{+}\case{1}{2}) \,\,, 
\\ \label{H_EL-PH}
H_{\rm el-ph} &=& \sum_{mn{\bf Q}} \hbar \omega_{\bf Q} 
g_{{\bf Q}mn}  a_m^{\dag} a_n (b_{\bf Q}^{\dag}{+}b_{-{\bf Q}}) \,\,,
\end{eqnarray}
describe the electrons, the phonons, and the electron-phonon interaction, 
respectively. Here, the operators $a_m^{(\dag)}$ and 
$b_{\bf Q}^{(\dag)}:=b_{{\bf q} \lambda}^{(\dag)}$ 
annihilate (create) an electron at equivalent lattice sites ${\bf R}_m=
\{ {\bf R}_1,\dots,{\bf R}_N \}$ with energy 
$\varepsilon_{mm}$ and a phonon in the mode $\lambda$ with wavevector 
${\bf q}$ and frequency $\omega_{\lambda {\bf q}}=:\omega_{\bf Q}$, 
respectively. The strength of the electron-lattice interaction in Eq.\ 
(\ref{H_EL-PH}) is determined by the dimensionless quantities 
$g_{{\bf Q}mn}$, which describe both the {\it local} coupling to the 
on-site energies $\varepsilon_{mm}$ (Holstein model) and the {\it nonlocal} 
coupling to the transfer integrals $\varepsilon_{mn}$ ($m \neq n$, 
Peierls model). The latter type of coupling will be shown to play a crucial 
role in the understanding of the anisotropic $T$ dependence of the mobilities.

For the description of the electrical conductivity, we use linear-response 
theory and  
introduce a conductivity tensor $\sigma_{\alpha \beta} (\omega)$ that 
relates the electric field ${\bf E}(\omega)$ present in the sample to the 
current ${\bf J}(\omega)$ that is induced by this field, $J_{\alpha}(\omega) 
=\sum_{\beta} \sigma_{\alpha \beta} (\omega) E_{\beta}(\omega)$.
Here, we concentrate on the mobilities $\mu_{\alpha}$ which 
are obtained in the zero-frequency limit according to 
$\sigma_{\alpha \alpha} (\omega{\rightarrow}0) = N_c e_0 \mu_{\alpha}$, 
where $e_0$ equals the negative (positive) elementary charge in case of 
electron (hole) conduction and $N_c$ denotes the number of charge carriers. 
In thermal equilibrium, the mobilities $\mu_{\alpha}$ can be calculated by 
means of the Kubo formula for electrical conductivity \cite{Mahan}, 
\begin{equation} \label{KUBO_FORMULA}
\mu_{\alpha} = \frac{1}{2 N_c e_0 k_B T} \int_{-\infty}^{+\infty} dt \, 
\left< j_{\alpha}(t) j_{\alpha}(0) \right> \,,
\end{equation}
which relates the mobility to a microscopic current-current 
correlation function. The r.h.s.\ of Eq.\ (\ref{KUBO_FORMULA}) 
contains the quantum-mechanical current operator  
${\bf j}(t)=e^{\frac{i}{\hbar}Ht}~{\bf j}~e^{-\frac{i}{\hbar}Ht}$
where ${\bf j}$ is obtained from the polarization operator 
${\bf P}= e_0 \sum_m {\bf R}_m a_m^{\dag} a_m $ via the Heisenberg 
equation of motion 
${\bf j}= \frac{d {\bf P}}{d t}= \frac{1}{i \hbar} [{\bf P}, H]$.
It is important to stress that the current ${\bf j}$ 
consists of two different contributions,
${\bf j}:={\bf j}^{(I)}+{\bf j}^{(II)}$, 
which are given by  
\begin{eqnarray} \label{DEF_J1}
{\bf j}^{(I)} &=& \frac{e_0}{i \hbar} \sum_{mn} 
({\bf R}_m{-}{\bf R}_n) \varepsilon_{mn} a_m^{\dag} a_n \,\,,\\ \label{DEF_J2}
{\bf j}^{(II)} &=& \frac{e_0}{i \hbar} \sum_{mn{\bf Q}} 
({\bf R}_m{-}{\bf R}_n) \hbar \omega_{\bf Q} g_{{\bf Q}mn} 
(b^{\dag}_{\bf Q}{+}b_{-{\bf Q}}) a_m^{\dag} a_n \,\,,
\end{eqnarray}
and originate from $H_{\rm el}$ and $H_{\rm el-ph}$, respectively. While the 
current ${\bf j}^{(I)}$ contains only electron operators, the second 
term ${\bf j}^{(II)}$ describes a phonon-assisted current that stems 
solely from the nonlocal 
electron-phonon coupling and has no counterpart in local-coupling theories. 

The evaluation of the Kubo formula ({\ref{KUBO_FORMULA}) is performed 
by the method of canonical transformation \cite{Mahan} 
within a generalized version that accounts for both local and nonlocal 
coupling \cite{Munn,Zhao,Hannewald}. By means of this technique, the 
electron-phonon interaction is incorporated nonperturbatively, in 
accordance with the underlying physical picture of small-polaron bands. 
Here, we skip the details of the quite tedious derivation and the 
approximations involved (which will 
be published elsewhere \cite{Hannewald2}) but rather concentrate on the 
final results for the polaron mobilities in the case of 
dispersionless optical phonons 
($\omega_{\bf Q} \rightarrow \omega_{\lambda}$) for which 
symmetry requirements impose the form $g_{{\bf Q}mn} = 
\frac{1}{2\sqrt{N}} g_{\lambda mn} (e^{-i {\bf q} \cdot 
{\bf R}_m}{+}e^{-i {\bf q} \cdot {\bf R}_n})$. 

It is instructive to consider first the 
mobility  $\mu_{\alpha}^{(I)}$ due to the currents 
${\bf j}^{(I)}$ in Eq.\ (\ref{DEF_J1}), which we obtain as    
\begin{equation} \label{RESULT_MOBIL_WO_PAC}
\mu_{\alpha}^{(I)} = 
\frac{e_0}{2 k_B T \hbar^2} \sum_{n \neq m} (R_{\alpha m}{-}R_{\alpha n})^2
 \int_{-\infty}^{+\infty} dt \,  \left[ \varepsilon_{mn} \, 
e^{-\sum_{\lambda} G_{\lambda} [1{+}2N_{\lambda}{-}\Phi_{\lambda}(t)]} 
\right]^2 e^{-\Gamma^2 t^2} \,\,,
\end{equation}
where $N_{\lambda}{=}\,(e^{\hbar \omega_{\lambda} / k_B T}{-}1)^{-1}$
denote the phonon occupation numbers. The extra factor involving the  
phenomenological broadening parameter $\Gamma$ accounts for scattering 
processes beyond the model, e.g., due to defects. Those terms in the 
exponent of Eq.\ 
(\ref{RESULT_MOBIL_WO_PAC}) that contain the time-dependent auxiliary 
function $\Phi_{\lambda}(t)= (1{+}N_{\lambda})e^{-i \omega_{\lambda}t}
+N_{\lambda}e^{i \omega_{\lambda}t}$ describe {\it incoherent} scattering 
events involving actual changes in phonon numbers (hopping) whereas the 
remaining terms account for purely {\it coherent} scattering processes 
(bandwidth narrowing). While our above result for $\mu_{\alpha}^{(I)}$ can 
be compared to the Holstein-model mobilities obtained by Kenkre {\it et al.} 
\cite{Kenkre,dispersion}, an important difference is that the 
exponent in our formula (\ref{RESULT_MOBIL_WO_PAC}) is not only governed 
by the 
local coupling but by effective coupling constants 
$G_{\lambda} = (g_{\lambda mm})^2 + \case{1}{2} \sum_{k \neq m} 
(g_{\lambda mk})^2$ that are composed of both the local {\it and} nonlocal 
ones.

So far, we have neglected the contributions from the phonon-assisted 
currents ${\bf j}^{(II)}$ of Eq.\ (\ref{DEF_J2}). If we include them, we  
obtain the total mobilities 
$\mu_{\alpha}$ within the Holstein-Peierls model, which can be expressed in 
a form identical 
to Eq.\ (\ref{RESULT_MOBIL_WO_PAC}) for $\mu_{\alpha}^{(I)}$, 
but with the replacement 
\begin{equation} \label{RESULT_MOBIL}
(\varepsilon_{mn})^2 \rightarrow 
(\varepsilon_{mn}{-}\Delta_{mn})^2 + \case{1}{2} \sum_{\lambda} 
(\hbar \omega_{\lambda} g_{\lambda mn})^2 \Phi_{\lambda}(t) \,.
\end{equation}
Here, we have introduced 
the shorthand notation  $\Delta_{mn}{=}\case{1}{2}\sum_{\lambda} \hbar 
\omega_{\lambda} \left[g_{\lambda mn} (g_{\lambda mm}{+}g_{\lambda nn})
+\frac{1}{2} \sum_{k \neq m,n}g_{\lambda mk} g_{\lambda kn}\right]$.
An essential qualitative difference between 
$\mu_{\alpha}^{(I)}$ and $\mu_{\alpha}$
is that only the total mobilities $\mu_{\alpha}$ can account for an 
anisotropic $T$ dependence. This is due to the additional terms containing 
$\Phi_{\lambda}(t)$ in Eq.\ (\ref{RESULT_MOBIL}), which describe 
incoherent scattering processes that are absent in local-coupling 
theories and reflect an inherently transport-promoting 
effect solely caused by the nonlocal coupling. The strength of the 
resulting phonon-assisted hopping and, hence, the anisotropy in the 
$T$ dependence is basically determined by the ratios 
$\hbar \omega_{\lambda} g_{\lambda mn}/ \varepsilon_{mn}$, 
which may be dramatically different for different directions.

As a first application of the above theory, we perform model studies for 
naphthalene, which crystallizes in a monoclinic structure ($P2_{1/a}$) 
and exhibits a herringbone stacking with two equivalent molecules per unit 
cell. In order to obtain the material-specific parameters 
${\bf R}_m$, $\varepsilon_{mn}$, $g_{\lambda mn}$, and $\omega_{\lambda}$ 
we use the following 3-step strategy. 

First, we determine 
the equilibrium structure of the crystal by means of state-of-the-art DFT-LDA 
calculations using the {\it ab initio} total-energy and molecular dynamics 
program VASP \cite{VASP}. For 
the lattice parameters we obtain the values $a=7.68\,{\rm \AA}$, 
$b=5.76\,{\rm \AA}$, $c=8.35\,{\rm \AA}$, and the monoclinic angle 
$\beta=125.7^{\rm o}$. For 
this geometry, the {\it inter}molecular optical-phonon energies 
($\hbar \omega_1 = 10.7 \, {\rm meV}$, $\hbar \omega_2 = 14.2 \, {\rm meV}$, 
$\hbar \omega_3 = 17.4 \, {\rm meV}$) and polarizations 
${\bf e}_{\lambda}$ are obtained within the rigid-molecule approximation, 
using a doubled Brillouin zone corresponding to the lattice $\{{\bf R}_m\}$
of all the molecules \cite{rotate}. In principle, our approach does 
also allow the inclusion 
of {\it intra}molecular phonons. However, in a recent Letter on oligo-acene 
molecules \cite{Coropceanu} it was reported that those intramolecular phonons 
that couple most strongly to the electrons have significantly higher 
frequencies than the intermolecular phonons considered here. Hence, while the 
zero-point fluctuations of the intramolecular phonons will lead to an 
isotropic overall reduction of the mobility values \cite{reduction}, their 
influence on 
the actual $T$ dependence will be small due to the low occupation of 
these modes up to room temperature.  

In the second step, we obtain the values $\varepsilon_{mn}$
from a fit of the ground-state {\it ab initio} HOMO and LUMO energy bands to a 
tight-binding model, including the on-site energy and the six most 
important transfer integrals between nearest neighbors, i.e., 
$\{mn\}{=}\{0,a,b,c,ac,ab,abc\}$ belonging to 
${\bf R}_m{-}{\bf R}_n={\bf 0},{\pm}\bbox{a},{\pm}\bbox{b},{\pm}\bbox{c},
{\pm}(\bbox{a}{+}\bbox{c}),{\pm}(\case{\bbox{a}}{2}{\pm}\case{\bbox{b}}{2})$, 
and 
${\pm}(\case{\bbox{a}}{2}{\pm}\case{\bbox{b}}{2}{+}\bbox{c})$, respectively.
 
In the third step, we rotate the molecules by amplitudes 
$\Delta u_{\lambda}$ according to 
the polarizations ${\bf e}_{\lambda}$ of phonon mode $\lambda$ 
(``frozen phonon'') and fit 
the new resulting {\it ab initio} bandstructure again to the 
tight-binding model. Then, the electron-phonon coupling constants 
$g_{\lambda mn}$ are obtained from the changes in the transfer 
integrals by numerical differentiation $g_{\lambda mn}{=}\case{1}{\hbar 
\omega_{\lambda}} \case{\Delta \varepsilon_{mn}}{\Delta u_{\lambda}}$, in the 
limit $\Delta u_{\lambda} \rightarrow 0$. 
All calculated parameters are compiled in Table \ref{Tab1}. Note that our 
{\it calculated} effective coupling constant 
$(g_{\rm eff})^2{:=}\sum_{\lambda =1}^3 G_{\lambda}{=}(1.52)^2$ for the 
LUMO band is comparable in value to the directionally-dependent 
{\it fitted} coupling 
values of $1.62$, $1.83$, and $1.88$, as obtained by Kenkre {\it et al.} 
\cite{Kenkre}.
Finally, for the line broadening, we have chosen a 
small value of $\hbar \Gamma = 0.1 \,{\rm meV}$, corresponding to the 
case of ultrapure crystals \cite{broadening}.

In Fig.\ \ref{fig1}, we present the electron and hole mobilities in 
naphthalene crystals as a function of temperature $T$ for the $a$, $b$, 
and $c'$ directions, with $c'$ being perpendicular to the $ab$-plane of the 
molecular layers. We make three important observations. 
First, the hole mobilities are generally larger 
than the corresponding electron mobilities. Only in the $a$ direction do we 
observe comparable mobilities for electrons and holes at 
$T \approx 300 \, {\rm K}$. This general trend is in good agreement 
with experiments of Karl and co-workers \cite{Karl,Warta}.  Importantly, the 
lower electron mobilities cannot be understood by simply looking at the 
transfer integrals of Table \ref{Tab1} but they are mainly due to their 
stronger coupling to the phonons (LUMO: $g_{\rm eff}=1.52$, 
HOMO: $g_{\rm eff}=0.92$). 

Second, while the hole 
mobilities in the different directions decrease very similarly with $T$, 
the electron mobilities exhibit a pronounced anisotropic $T$ dependence. In 
particular, we find a metallic (bandlike) behavior within the $ab$-plane 
and a slightly activated (hopping) transport in the $c'$ direction. Again, 
these findings from our 
{\it ab initio} calculations are in good agreement with experimental data 
\cite{Schein,Karl}, which show such a nearly $T$-independent mobility 
around $T \approx 150 \, {\rm K}$ exclusively in the $c'$ direction.  
As discussed above, this pronounced anisotropy is due to the phonon-assisted 
currents ${\bf j}^{(II)}$ and depends on the ratios 
$\hbar \omega_{\lambda} g_{\lambda mn} / \varepsilon_{mn}$. For the electrons, 
this ratio is particularly high for $\hbar \omega_2 g_{2c} / \varepsilon_c$,
which explains the peculiar flattening of $\mu_{c'}$ that is not seen at all
for $\mu_{c'}^{(I)}$ alone. 
Thus, our findings show clearly that anisotropy effects in the $T$ dependence 
can be explained in a natural way by the inclusion of nonlocal 
electron-phonon coupling, and do not require the use of additional concepts 
such as a directionally-dependent local couplings 
\cite{Kenkre} or correlated successive hops of 
strongly localized charge carriers \cite{Emin2}. 

Third, our calculated hole mobilities obey at elevated temperatures 
the power law $\mu_{\alpha} \propto T^{-2.5}$. This agrees nicely with 
Karl's experiments \cite{Karl} who found $\mu_{\alpha} \propto T^{-\gamma}$ 
($\gamma=2.9$, $2.5$, $2.8$). 
This peculiar power-law dependence, which has been poorly understood so far 
and for which several explanations have been put forward \cite{Warta}, can 
be straightforwardly understood from our theory. In the limit of small 
broadening $\hbar \Gamma$ and for $\mu_{\alpha} \approx \mu_{\alpha}^{(I)}$ 
(here fulfilled for the holes), a spectral analysis of Eq.\ 
(\ref{RESULT_MOBIL_WO_PAC}) shows that the dominant electron-phonon 
scattering processes are those that allow only 
energy exchange within each phonon mode but not between different modes, and 
the mobility can be expressed in terms of modified Bessel functions,
$\mu_{\alpha}^{(I)} \propto 
\frac{1}{\Gamma} \frac{1}{k_B T} \left( \prod_{\lambda =1}^3 
e^{-2 G_{\lambda} (1{+}2N_{\lambda})} I_0(z_{\lambda}) \right)$, 
where $z_{\lambda}=4 G_{\lambda} \sqrt{N_{\lambda}(1{+}N_{\lambda})}$. The 
limit of high $T$ is obtained using $I_0(z_{\lambda}{\rightarrow}\infty) 
\approx e^{z_{\lambda}} / \sqrt{2 \pi z_{\lambda}}$ and we find 
$\mu_{\alpha}^{(I)} \propto T^{-1} \left( \prod_{\lambda =1}^3 
T^{-0.5} \right) = T^{-2.5}$ which indicates that the exponent is actually a 
measure 
for the number of relevant phonon modes per molecule. For the electrons, the 
situation is more complicated due to a subtle interplay between 
the now significant contributions of ${\bf j}^{(II)}$ and 
the larger $g_{\rm eff}$ value. This prohibits an easy analytical 
evaluation but, by looking at our calculated in-plane electron mobilities in 
Fig.\ \ref{fig1}, we find at high $T$ an approximate $T^{-1.5}$ dependence, 
again in close agreement with the experimental observations \cite{Karl,Warta}.

In summary, we have presented a novel theoretical description of 
charge-carrier mobilities in ultrapure organic crystals based on 
a mixed Holstein-Peierls model where not only the local but, importantly, 
also the nonlocal electron-phonon coupling is included. The predictive power 
of our theory and the usefulness of the small-polaron band concept have been 
demonstrated by {\it ab initio} studies for naphthalene crystals where our 
results show a remarkably good agreement with experiments and provide new 
microscopic insight into several hitherto poorly understood phenomena
including the different behavior of electrons and holes as well as the 
peculiar algebraic and anisotropic temperature dependences.
Finally, since our theory may also be applied to 
other weakly-bonded materials with strong electron-phonon coupling, 
it may become an important contribution to the understanding of the 
intrinsic charge-carrier transport in these materials, too. 

We would like to thank D. Emin, S. Mazumdar, J. Pflaum, S. Stafstr\"om, 
J. van den Brink, and J. Zaanen for valuable 
discussions. Financial support by the Dutch Foundation for Fundamental 
Research on Matter (FOM) is gratefully acknowledged.

\begin{table}[t!]
\begin{center}
\begin{tabular}{|c|rrrr|rrrr| }\hline
                 &\multicolumn{4}{c|}{HOMO band~~~~}
                 &\multicolumn{4}{c|}{LUMO band~~~~}\\ \hline
    $\{mn\}$&$\varepsilon_{mn}$&$g_{1mn}$&$g_{2mn}$&$g_{3mn}$
            &$\varepsilon_{mn}$&$g_{1mn}$&$g_{2mn}$&$g_{3mn}$\\ 
            &(meV)&&&&(meV)&&&\\ \hline
$0$&&$-0.04$&$0.33$&$0.33$&&$-0.08$&$0.09$&$0.05$ \\
$a$&$-29$&$-0.03$&$-0.24$&$0.01$&$1$&$0.03$&$-0.05$&$0.01$ \\
$b$&$-59$&$0.43$&$0.05$&$0.20$&$30$&$-0.87$&$0.09$&$0.00$ \\
$c$&$4$&$0.01$&$0.09$&$0.02$&$1$&$-0.15$&$0.53$&$-0.05$ \\
$ac$&$6$&$-0.02$&$0.00$&$-0.05$&$-3$&$0.07$&$-0.18$&$-0.01$ \\
$ab$&$17$&$-0.25$&$-0.25$&$0.05$&$-72$&$0.11$&$-0.69$&$-0.11$ \\
$abc$&$-24$&$0.15$&$0.08$&$-0.06$&$-4$&$-0.08$&$0.28$&$-0.08$ \\ \hline \hline
   &$g_{\rm eff}$&$G_1$&$G_2$&$G_3$&$g_{\rm eff}$&$G_1$&$G_2$&$G_3$ \\ \hline
   &$0.92$&$0.36$&$0.32$&$0.16$&$1.52$&$0.83$&$1.44$&$0.04$ \\ \hline
   \end{tabular}
\end{center}
\caption{Top: transfer integrals $\varepsilon_{mn}$ and electron-phonon 
coupling constants $g_{\lambda mn}$ for the HOMO and LUMO bands 
of naphthalene crystals. Bottom: effective coupling constants 
$(g_{\rm eff})^2{:=}\sum_{\lambda =1}^3 G_{\lambda}$, where 
$G_{\lambda} = g^2_{\lambda 0}{+}g^2_{\lambda a}{+}
g^2_{\lambda b}{+}g^2_{\lambda c}{+}g^2_{\lambda ac}{+}2g^2_{\lambda ab}{+}
2g^2_{\lambda abc}$.}
\label{Tab1}
\end{table}

\begin{figure}
\caption{Charge-carrier mobilities vs temperature $T$ for 
the $a$, $b$, and $c'$ directions in naphthalene crystals. Top 3 curves: 
$\mu_{\alpha}$ for holes, middle 3 curves: $\mu_{\alpha}$ for electrons, 
bottom curve: $\mu_{c'}^{(I)}$ for electrons. As a guide to the eye, several 
power laws $\mu_{\alpha} \propto T^{-\gamma}$ are also depicted.
All mobilities are plotted in units of $\frac{{\rm cm}^2}{\rm V s}
\times \frac{76 \, {\rm meV}}{\hbar \Gamma}$ but an unknown prefactor $r$ 
\protect\cite{reduction} should be included.} 
\label{fig1}
\end{figure}


\begin{references}
\bibitem{dispersion} The mobility expressions in Ref. \cite{Kenkre} 
contain also the phonon bandwidths, which are, however, very small and 
neglected by us.
\bibitem{rotate} In this approach, the three optical phonon modes 
involved are purely rotational at the Brillouin-zone center.
\bibitem{reduction} From the Huang-Rhys factors $S_{\lambda}$ of 
Ref.\ \cite{Coropceanu}, 
we estimate that the intramolecular phonons reduce the mobilities by a 
factor $r < \left(e^{-\sum_{\lambda} S_{\lambda}} \right)^2 \approx 0.34$.
\bibitem{broadening} We have carefully checked that, for the temperature 
region shown in Fig.\ \ref{fig1}, smaller values of $\hbar \Gamma$ do not 
change the actual $T$ dependence of the mobilities but lead only to a 
scaling $\propto 1/\Gamma$ of their overall values.
\end{references}
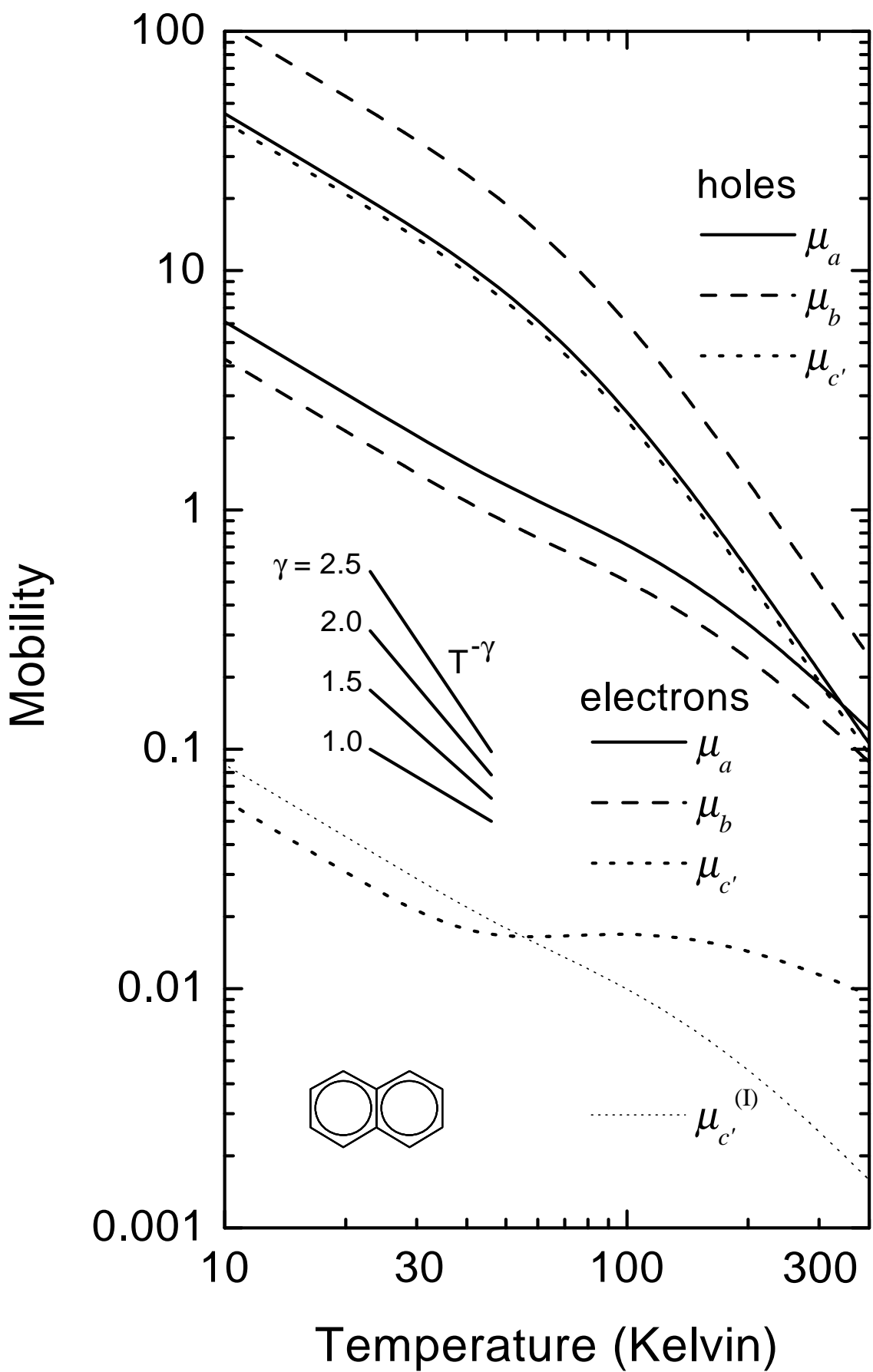
\end{document}